\documentclass[article, twocolumn,
   aps, pra,
  amsmath,amssymb,
  longbibliography,
  ]{revtex4-1}
\usepackage{graphicx,color}
\usepackage{amsmath}
\usepackage{natbib}
\usepackage{epsfig}
\begin{document}

\title{Bridgman formula for the thermal conductivity of atomic and molecular liquids}

\author{S. A. Khrapak}\email{Sergey.Khrapak@gmx.de}
\affiliation{Joint Institute for High Temperatures, Russian Academy of Sciences, 125412 Moscow, Russia}

\begin{abstract}
A simple and popular Bridgman's model predicts a linear correlation between the thermal conductivity coefficient and the sound velocity of dense liquids. A proportionality coefficient proposed originally is fixed and independent of the liquid molecular structure. This work reports a systematic analysis of correlations between thermal conductivity  and sound velocity in simple model systems (hard sphere and Lennard-Jones fluids), monoatomic liquids (argon and krypton), diatomic liquids (nitrogen and oxygen), and several polyatomic liquids (water, carbon dioxide, methane, and ethane). It is demonstrated that linear correlations are well reproduced for model fluids as well as real monoatomic and diatomic liquids, but seem less convincing in polyatomic molecular liquids. The coefficient of proportionality is not fixed; it is about unity for monoatomic liquids and generally increases with molecular complexity. Some implications for the possibility to predict the thermal conductivity coefficients are discussed.              
\end{abstract}

\date{\today}

\maketitle

\section{Introduction}

About a century ago Bridgman proposed a formula, which relates the coefficient of thermal conductivity with the sound velocity and density of dense liquids~\cite{Bridgman1923}:
\begin{equation}\label{tc1}
\lambda = \frac{2c_s}{\Delta^{2}},
\end{equation}
where $\lambda$ is the coefficient of thermal conductivity, $c_s$ is the sound velocity, and $\Delta$ is the mean separation between the molecules related to the liquid number density $\rho$ via $\Delta=\rho^{-1/3}$. Note that the thermal conductivity coefficient is defined here via Fourier's law as follows $q=\lambda\nabla (k_{\rm B}T)$, so that the Boltzmann constant $k_{\rm B}$ does not appear in the definition of $\lambda$ (hence $\lambda$ is measured in cm$^{-1}$s$^{-1}$). Bridgman
proposed the following very simple physical picture to deduce formula (\ref{tc1}). Liquid structure was approximated by a cubic lattice with the lattice constant $\Delta$. Bridgman assumed that each molecule has an energy $2k_{\rm B}T$ (half potential and half kinetic, he noted).  
If a temperature gradient $dT/dx$ is present, the energy difference between neighbouring molecules in the direction of the temperature gradient is $2k_{\rm B}\Delta dT/dx$. He finally assumed that this energy difference is transferred at the speed of sound and this corresponds to the energy flux through the characteristic area controlled by each liquid molecule, which is $\Delta^2$. This immediately leads us to Eq.~(\ref{tc1}).

Bridgman's model can be criticised on several grounds. Most obvious are the assumption of the cubic lattice and the use of a fixed energy $2k_{\rm B}T$ per molecule (in a more appropriate consideration specific heat would have to appear from the partial derivative of energy with respect to temperature). On the other hand, Bridgman's expression was apparently the first theoretical formula applied to the subject. It demonstrated more than just an agreement of order of magnitude when compared to experimental values for eleven liquids considered originally in Ref.~\cite{Bridgman1923} (even though an incorrect value for the Boltzmann's constant was chosen in the original work~\cite{Bridgman1923,ZhaoJAP2021}). Since our understanding of dynamics and transport in the liquid state remains far from complete, despite considerable efforts in this direction~\cite{FrenkelBook,McLaughlin1964,BarkerRMP1976,GrootBook,BalucaniBook,
MarchBook,HansenBook}, Bridgman's model still represents a useful simple reference model and is often discussed in contemporary literature~\cite{ZhaoJAP2021,XiCPL2020,KhrapakPRE01_2021,Chen2021}. 

From a general perspective, Bridgman postulated that the energy is transferred at the speed of sound, which leads to a linear correlation between the thermal conductivity coefficient and the sound velocity as in Eq.~(\ref{tc1}). Similar correlation can be obtained within a vibrational model of heat transfer in simple fluids with soft interactions under the assumption that fluid supports only sound-like longitudinal mode (thus neglecting the transverse modes that are also supported in dense fluids~\cite{OhtaPRL2000,HosokawaJPCM2015,YangPRL2017,BrykJCP2017,KhrapakJCP2019,
KryuchkovSciRep2019}). The exact numerical coefficient remains somewhat contradictory, values between 2 and 3 were used in the literature~\cite{ZhaoJAP2021,XiCPL2020,KhrapakPRE01_2021,BirdBook}.

The purpose of this paper is to report a systematic analysis of correlations between the thermal conductivity coefficient and the sound velocity in liquids with different molecular composition. The idea is that the coefficient of proportionality in Eq.~(\ref{tc1}) is likely to be affected by the complexity of molecular structure, the point which seemingly has been overlooked in previous studies. We will see that this is indeed correct. The analysis will begin with two important models -- hard sphere and Lennard-Jones fluids in Sec.~\ref{Models}. Then two liquefied noble gases, argon and krypton, as examples of monoatomic liquids will be examined in Sec.~\ref{Nobles}. Liquids composed of diatomic molecules (represented by nitrogen and oxygen), triatomic molecules (represented by water and carbon dioxide) and polyatomic molecules (represented by methane and ethane) will be further considered in Sec.~\ref{Molecules}. This will be followed by discussion and conclusion in Sec.~\ref{Conclusion}.         

\section{Hard sphere and Lennard-Jones fluids}\label{Models}

We start with two model fluids, which have been playing exceptional role in condensed matter research. The first is the hard sphere (HS) system, which is characterized by extremely hard- and short-ranged repulsive interaction. The interaction energy is infinite for overlapping spheres and is zero otherwise: 
\begin{equation}
    \phi(r)= 
\begin{cases}
    \infty,&  r< \sigma\\
    0,         & r\geq \sigma,
\end{cases}
\end{equation}
where $\sigma$ is the sphere diameter and $r$ is the distance between the centres of two spheres. The reduced number density $\rho^{*}=\rho\sigma^3$ is the only state variable in the case of the HS system (though the pressure, for instance, is also proportional to the temperature). 


The second is the Lennard-Jones (LJ) system, which is often used to approximate interactions in noble gases. The LJ potential reads 
\begin{equation}
\phi(r)=4\epsilon\left[\left(\frac{\sigma}{r}\right)^{12}-\left(\frac{\sigma}{r}\right)^{6}\right], 
\end{equation}
where  $\epsilon$ and $\sigma$ are the energy and length scales (or LJ units), respectively. The dynamics and thermodynamics are determined by two state variables, usually by thereduced number density $\rho^*=\rho\sigma^3$ and temperature $T^*=k_{\rm B}T/\epsilon$.

Both HS and LJ systems have been extensively studied in the context of condensed mater physics. Many results on transport properties of these systems have been published over the years. Here we use recent accurate molecular dynamics (MD) simulation results for the thermal conductivity coefficient of HS fluids reported by Pieprzyk {\it et al}.~\cite{Pieprzyk2020}. For the LJ fluid we 
make use of recently proposed reference correlations between the
transport coefficients of the LJ fluid and excess entropy~\cite{BellJPCB2019}, based on the celebrated Rosenfeld's excess entropy scaling~\cite{RosenfeldPRA1977,DyreJCP2018}. Practical expressions from this approach have a very simple functional form and allow for the reproduction of the most accurate simulation data to within
nearly their statistical uncertainty~\cite{BellJPCB2019}.
We chose the supercritical isotherm $T^*=2$ to minimize effects associated with critical enhancement. A recent observation that each of the reduced transport coefficients of self-diffusion, shear viscosity, and thermal conductivity of dense LJ fluids nearly coincides along different isotherms (subcritical and supercritical) when plotted as a function of density divided by its value at the freezing point~\cite{KhrapakPRE04_2021,KhrapakJPCL2022}, makes it sufficient to consider a single isotherm. 

To plot the thermal conductivity coefficients as functions of the sound velocity, the latter should be evaluated, too. 
The sound velocity is defined by the standard thermodynamics relation $c_{\rm s}=\sqrt{\left(\partial P/\partial \rho_m\right)_S}$, where $P$ is the pressure, $\rho_m=m\rho$ is the mass density, and the derivative is taken at constant entropy $S$. Simple expressions relating the sound velocity and the compressibility factor of the HS system can be found e.g. in Refs.~\cite{RosenfeldJPCM1999,KhrapakJCP2016}. The Carnahan-Starling equation of state (EoS)~\cite{CarnahanJCP1969} is then used. For the LJ fluid the EoS by Thol {\it et al.}~\cite{Thol2016} is employed in calculations. It is convenient to express the sound velocity in units of the characteristic thermal velocity $v_{\rm T}=\sqrt{k_{\rm B}T/m}$, where $m$ is the atomic or molecular mass. 

\begin{figure}
\includegraphics[width=8cm]{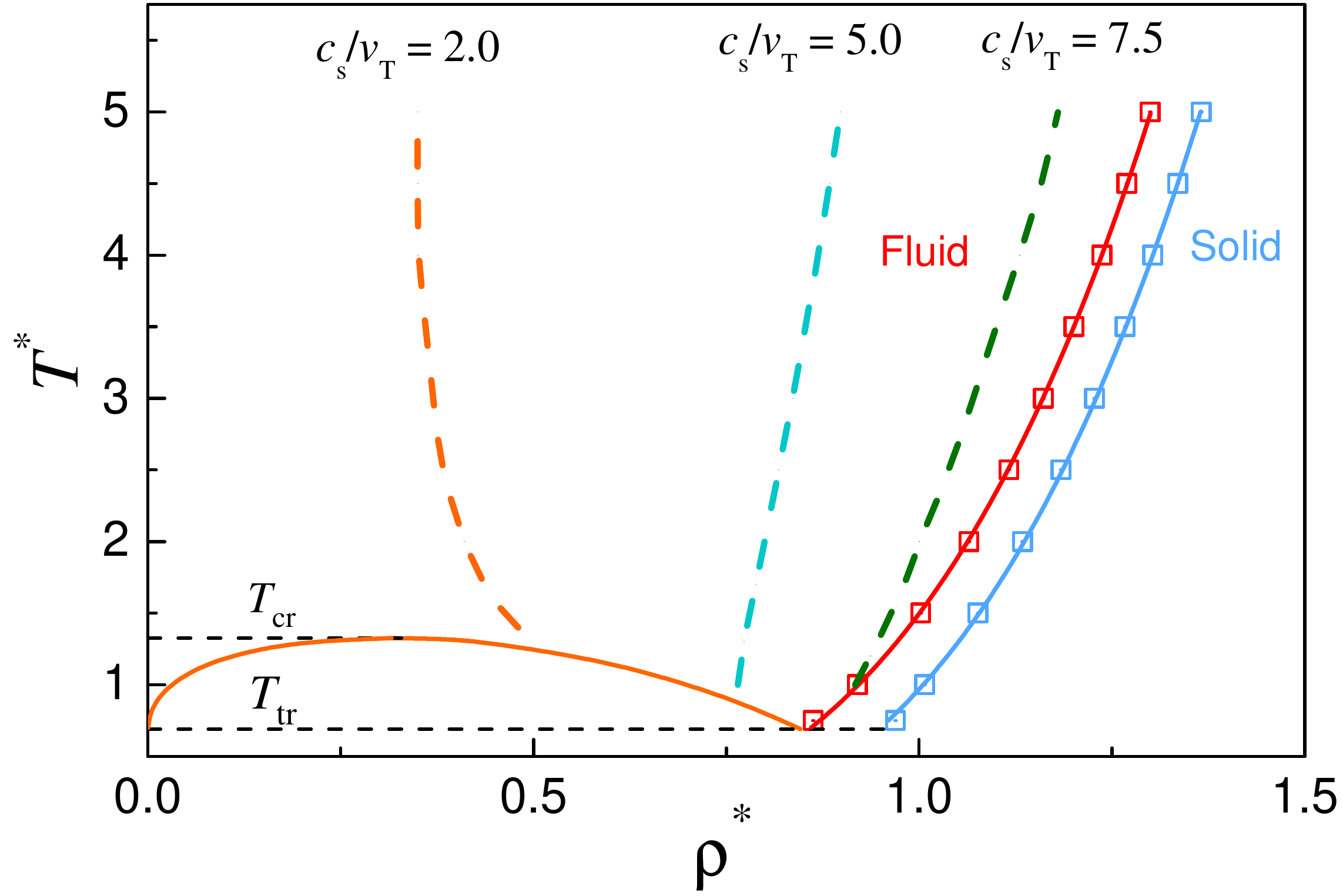}
\caption{(Color online) Phase diagram of the Lennard-Jones system. The squares correspond to the fluid-solid coexistence boundaries as tabulated in Ref.~\cite{SousaJCP2012}; the corresponding curves are simple fits of the form $T_{\rm fr}^*=2.111(\rho^*)^4-0.615(\rho^*)^2$ for the freezing curve and  $T_{\rm fr}^*=1.988(\rho^*)^4-1.019(\rho^*)^2$ for the melting curve.  The liquid-vapour boundary is plotted using the formulas provided in Ref.~\cite{HeyesJCP2019} with the reduced triple point and critical temperatures being $T_{\rm tr}\simeq 0.694$~\cite{SousaJCP2012} and $T_{\rm cr}\simeq 1.326$~\cite{HeyesJCP2019}. 
Three dashed curves correspond to constant values of the sound velocity divided by the thermal velocity, $c_{\rm s}/v_{\rm T}=2$, $5$, and $7.5$ (from left to right), as evaluated from the LJ equation of state proposed by Thol {\it et al}.~\cite{Thol2016}.}
\label{Fig0}
\end{figure}

Figure~\ref{Fig0} shows the phase diagram of the LJ system along with the three lines of constant ratio $c_{\rm s}/v_{\rm T}=2, 5$ and $7.5$. A couple of relevant observations should be pointed out. First of all, the curves of constant reduced sound velocity are not even approximately parallel to the freezing and melting curves in contrast to the curves of constant reduced transport coefficients. Therefore, the freezing density scaling of reduced transport coefficients~\cite{KhrapakPRE04_2021,KhrapakJPCL2022,KhrapakJCP2022_1} is not applicable to the sound velocity. Thus, correlations between the thermal conductivity and the sound velocity are different from trivial correlations between the two properties exhibiting a similar scaling behaviour~\cite{KhrapakJETPLett2021}. Second, and not unrelated to the first point, the reduced sound velocity varies considerably along the LJ fluid freezing curve.   It has the value $c_{\rm s}/v_{\rm T}\simeq 6.6$ at the triple point and increases to $c_{\rm s}/v_{\rm T}\simeq 8.6$ at $T^*=5$. Therefore, sound velocity does not belong to such freezing point quasi-invariants as the radial distribution function, structure factor and some reduced transport properties~\cite{HansenMolPhys1973,SaijaJCP2000,SaijaJCP2001,SaijaJCP2006,
CostigliolaPCCP2016}. It should also be noted that the observed relative variation of the thermodynamic sound velocity exceeds considerably the relative variations of the instantaneous longitudinal and transverse sound velocities (related to the instantaneous longitudinal and shear moduli) along the freezing curve of the LJ fluid~\cite{KhrapakMolecules2020}.

For the thermal conductivity coefficient properly reduced units should also be used. Throughout this paper the reduced thermal conductivity is expressed as:        
\begin{equation}\label{Rosenfeld}
\lambda_{\rm R}=\lambda\frac{\rho^{-2/3}}{v_{\rm T}}.
\end{equation}
This normalization is essential in the Rosenfeld's excess entropy scaling approach~\cite{RosenfeldPRA1977,RosenfeldJPCM1999_HS}, hence the subscript ${\rm R}$ is often used. We keep this tradition here. This normalization applies not only to the HS and LJ fluids considered in this section, but to all the fluids analyzed throughout the paper.

\begin{figure}
\includegraphics[width=8cm]{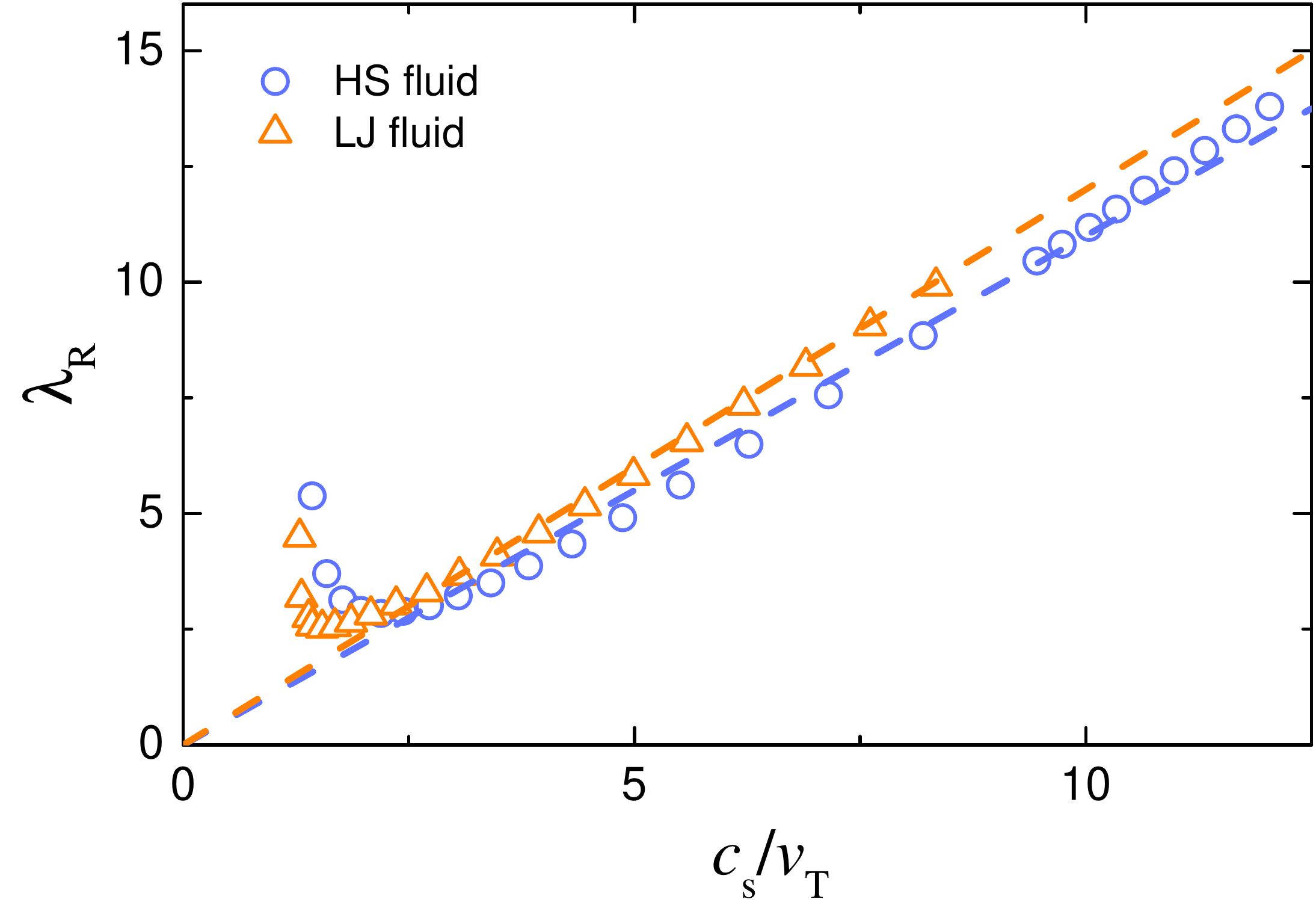}
\caption{(Color online) Macroscopically reduced thermal conductivity coefficient of HS and LJ fluids as a function of the reduced sound velocity. Circles correspond to the numerical results for the HS fluid from Ref.~\cite{Pieprzyk2020}. Triangles denote a modified entropy scaling approach of Ref.~\cite{BellJPCB2019} for the LJ fluid isotherm $T^*=2$. Dashed lines correspond to linear correlations suggested by Bridgman's formula (with the proportionality coefficient $1.1$ for HS and $1.2$ for LJ)}.
\label{Fig1}
\end{figure}

The main results of this section are presented in Figure~\ref{Fig1}. Here the reduced thermal conductivity coefficient $\lambda_{\rm R}$ is plotted versus the reduced sound speed $c_{\rm s}/v_{\rm T}$ for HS and LJ fluids. 
The two well defined branches are observed. In the dilute gaseous regime, the thermal conductivity coefficient can be estimated as
$\lambda\sim c_{\rm p} v_{\rm T} \rho \ell$, where $c_{\rm p}$ is the specific heat at constant pressure and $\ell$ is the mean free path between collisions~\cite{LifshitzKinetics}. In reduced units this yields $\lambda_{\rm R}\sim c_{\rm p}\rho^{1/3}\ell$. Taking into account that $\ell\sim 1/\rho\Sigma$, where $\Sigma$ is the collisional cross section, we see immediately that $\lambda_{\rm R}$ diverges as $\propto \rho^{-2/3}$ in the dilute gas limit. The sound veocity, on the other hand, approaches the dilute gas asymptote $c_{\rm s}=\sqrt{\gamma k_{\rm B}T/m}$, where $\gamma=c_{\rm p}/c_{\rm v}$ is the adiabatic index ($\gamma = 5/3$ for atomic substances with three degrees of freedom considered here). This corresponds to the data displayed to the left in Figure~\ref{Fig1}. The rest of the data correspond to the fluid regime. These data demonstrate rather good linear correlation between the thermal conductivity coefficient and the sound velocity of the form of Bridgman's expression (\ref{tc1}), but with slightly different proportionality coefficients, $\simeq 1.1$ for HS fluid and $\sim 1.2$ for LJ fluid. The fact that Bridgman's formula works relatively well for the HS fluid, but with the proportionality coefficient about unity has been recently pointed out in Ref.~\cite{KhrapakApplSci2022}. 

The existence and the location of a demarcation line (often referred to as Frenkel line) between liquid-like and gas-like dynamics of supercritical fluids has been a topic of major interest in recent years~\cite{Simeoni2010,BrazhkinPRE2012,BrazhkinPRL2013,BellJCP2020,
KhrapakJCP2022}. The presented results are consistent with the gas-to-liquid dynamical crossover occurring at $c_{\rm s}/v_{\rm T}\gtrsim 3$. Note in this context that the condition $c_{\rm s}\simeq 2 v_{\rm T}$ for a qualitative change between liquid-like and gas-like dynamics was previously discussed in the literature (note that the definition $v_{\rm T}=\sqrt{3k_{\rm B}T/m}$ was used there)~\cite{BrazhkinPRE2012,BrazhkinUFN2012}. 

It is also observed in Figure~\ref{Fig1} that the minimum values of $\lambda_{\rm R}$ for HS and LJ fluids are close ($\simeq 2.5$). This is not very surprising in view of universal lower bounds on the energy and momentum diffusion in liquids discussed recently~\cite{TrachenkoPRB2021}. In fact, minima in the reduced thermal conductivity and viscosity coefficients arise due to the crossover between the gas-like and liquid-like regimes of energy and momentum transport. Purely classical arguments suffice to demonstrate that the magnitudes of the minima should be expected to be quasi-universal~\cite{KhrapakPoF2022}.

\section{Real monoatomic liquids}\label{Nobles}

We have analysed recommended data for the sound velocity and thermal conductivity coefficients of liquefied argon and krypton provided in the National Institute of Standards and Technology (NIST) Reference Fluid Thermodynamic and Transport Properties Database (REFPROP 10.0)~\cite{Refprop}. For each liquid, three supercritical and one subcritical isotherms have been selected. The maximum density usually (but not always due to limitations of models implemented in REFPROP 10.0) corresponds to the freezing density. The lowest density is that of the saturated liquid for the subcritical isotherm and usually about two orders of magnitude lower than the freezing density for supercritical isotherms. At such low densities gas-like mechanism of energy transfer operates and the thermal conductivity coefficient expressed using Rosenfeld's normalization diverges as $\rho^{-2/3}$. 

\begin{figure}
\includegraphics[width=8cm]{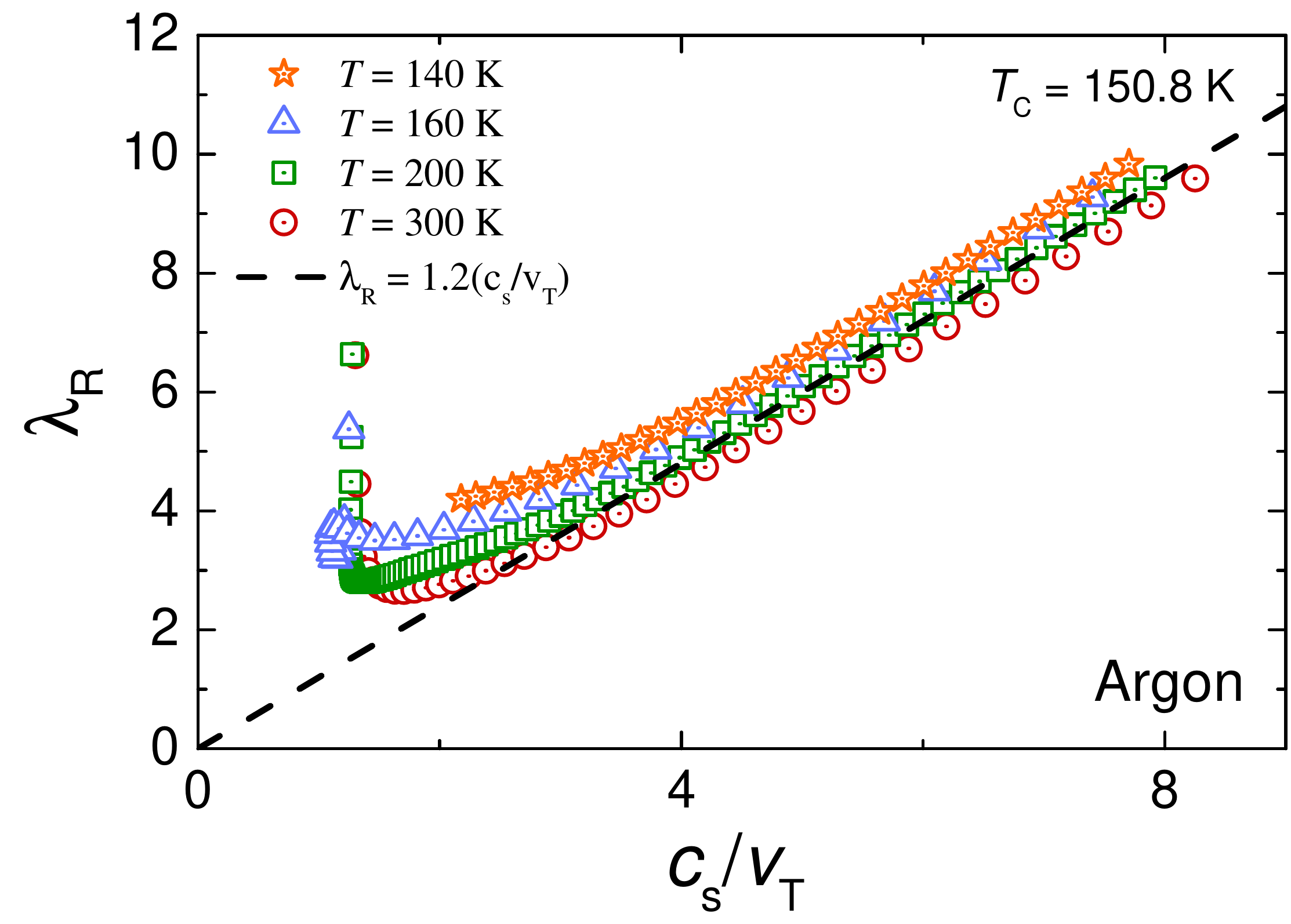}
\caption{(Color online) Macroscopically reduced thermal conductivity coefficient of liquid argon as a function of the reduced sound velocity. Symbols denote the values from REFPROP 10.0 along three supercritical and one subcritical isotherm (see label). The dashed line corresponds to linear correlations suggested by Bridgman's formula with the proportionality coefficient $1.2$.}
\label{Fig2}
\end{figure}

The EoS and the thermal conductivity model of argon, implemented in REFPROP are those from Refs.~\cite{Tegeler1999} and \cite{Lemmon2004}, respectively. For krypton, the EoS and thermal conductivity models are those from Refs.~\cite{Lemmon2006,Huber2018}. Note that the transport data for Krypton were either not available or were not correlated in REFPROP 10.0.  Thermal conductivity values were calculated with an extended corresponding states model in a predictive mode~\cite{Huber2018}.
 
\begin{figure}
\includegraphics[width=8cm]{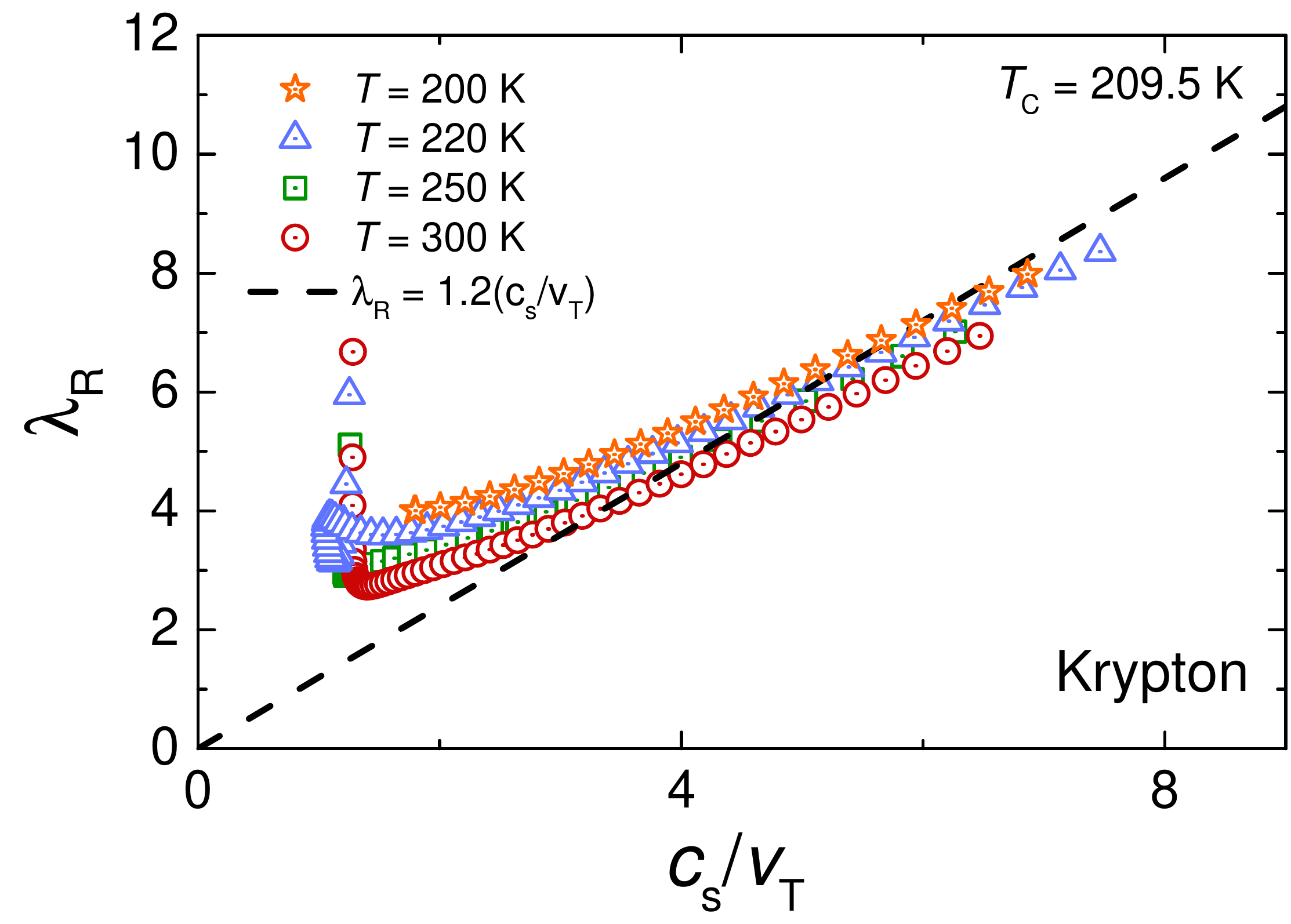}
\caption{(Color online) Macroscopically reduced thermal conductivity coefficient of liquid krypton as a function of the reduced sound velocity. Symbols denote the values from REFPROP 10.0 along three supercritical and one subcritical isotherm (see label). The dashed line corresponds to linear correlations suggested by Bridgman's formula with the proportionality coefficient $1.2$.}
\label{Fig3}
\end{figure}

The results are presented in Figs.~\ref{Fig2} and \ref{Fig3}. The linear correlation between the thermal conductivity coefficient and the sound velocity is observed at sufficiently large values of $c_{\rm s}/v_{\rm T}$ . For argon, the correlation of the form of Eq.~(\ref{tc1}) is appropriate. For krypton, a linear expression with the residual term 
\begin{equation}\label{lin}
\lambda_{\rm R}=\alpha(c_{\rm s}/v_{\rm T})+\beta,
\end{equation}
would be somewhat more appropriate, but has not been attempted. The proportionality coefficient between $\lambda_{\rm R}$ and $c_{\rm s}/v_{\rm T}$ is close to unity ($\simeq 1.2$), similarly to the case of LJ and HS fluids. The linear correlation holds for sound speeds from $c_{\rm s}/v_{\rm T}\gtrsim 3$, up to the freezing point. 

Other observations to be noted are:

(i) {\it Quasi-universality in the dependence of $\lambda_{\rm R}$ on $c_{\rm s}/v_{\rm T}$}: For each substance this dependence has a relatively weak temperature dependence in the considered range of temperatures (at the same time, entropy scaling and freezing density scaling are expected to result in a better collapse of experimental data~\cite{BellJPCB2019,KhrapakJCP2022_1}); \\

(ii) {\it Quasi-universality of the minima values}: The values of $\lambda_{\rm R}$ at the minima are almost the same, $\lambda_{\rm R}\simeq 3$ (except at near-critical temperatures, where they are higher due to critical enhancement). Classical arguments explaining why it should be so for simple model and real atomic fluids can be found in Ref.~\cite{KhrapakPoF2022};\\

(iii) {\it Quasi-universality of $\lambda_{\rm R}$ at freezing}:  The values of $\lambda_{\rm R}$ near the freezing point are almost the same, $\lambda_{\rm R}\simeq 10$ (note that not all data sets are terminating at the freezing point). The vibrational paradigm of energy transport in dense fluids explains this universality very well~\cite{KhrapakPoF2022,KhrapakJCP2022_1};  

Thus, Bridgman's formula remains applicable also in the case of real monoatomic liquids, as exemplified by liquefied argon and krypton here. Similar to simple model fluids, the coefficient of proportionality is about unity. In the next Section, we will focus on several bi-, tri-, and polyatomic liquids to continue our analysis.

\section{Molecular liquids}\label{Molecules}

\subsection{Diatomic molecules}

As examples of diatomic molecules let us consider liquid nitrogen and oxygen. Data are again taken from REFPROP 10.0. The EoS and the thermal conductivity models used in the database are those from Refs.~\cite{Span2000,Lemmon2004} for N$_2$ and Refs.~\cite{Schmidt1985,Lemmon2004} for O$_2$. We consider three supercritical and one subcritical isotherms for N$_2$ and two supercritical and two subcritical isotherms for O$_2$. 

\begin{figure}
\includegraphics[width=8cm]{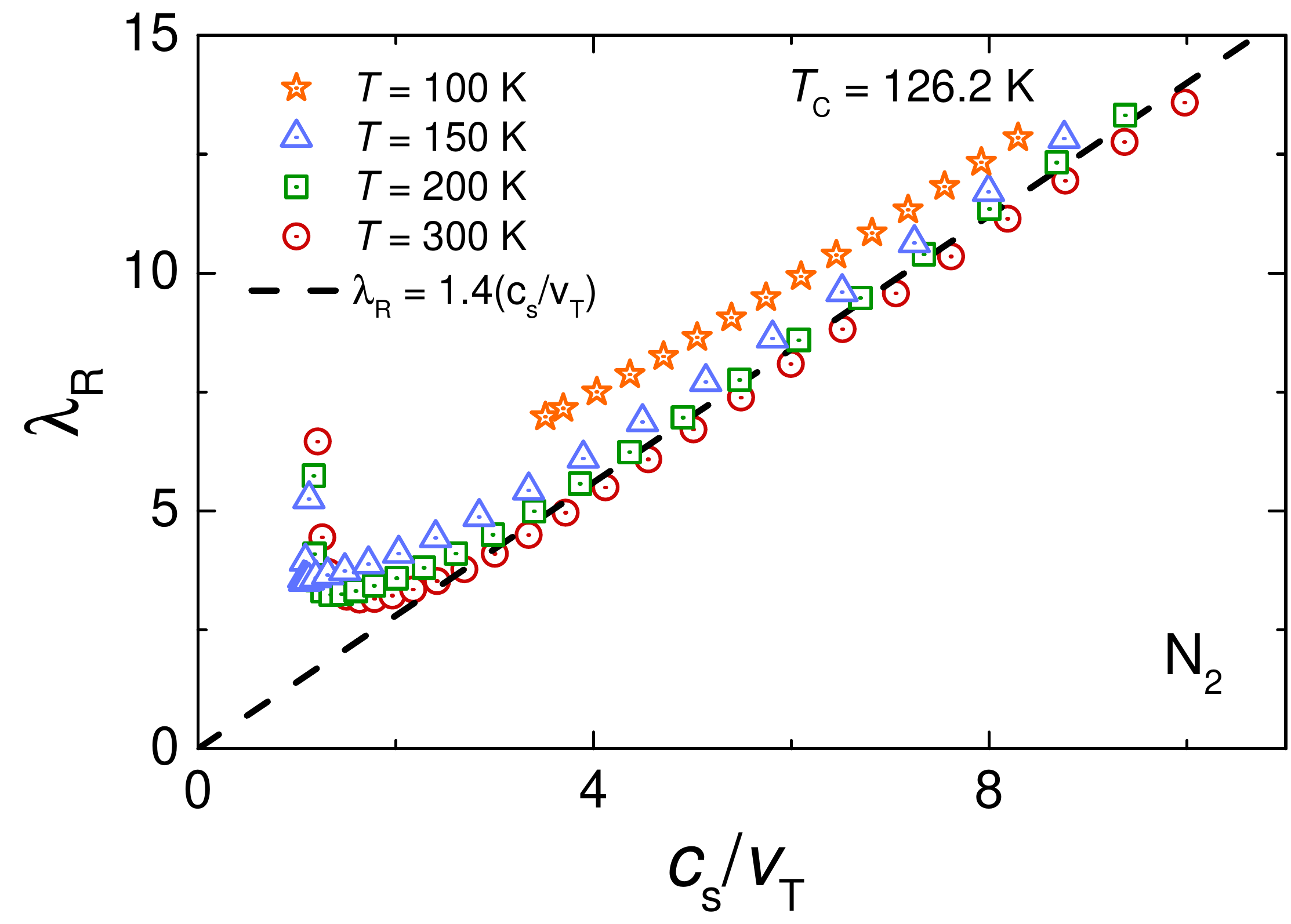}
\caption{(Color online) Macroscopically reduced thermal conductivity coefficient of liquid nitrogen as a function of the reduced sound velocity. Symbols denote the values from REFPROP 10.0 along three supercritical and one subcritical isotherm (see label). The dashed line corresponds to linear correlations suggested by Bridgman's formula with the proportionality coefficient $1.4$.}
\label{Fig4}
\end{figure}

\begin{figure}
\includegraphics[width=8cm]{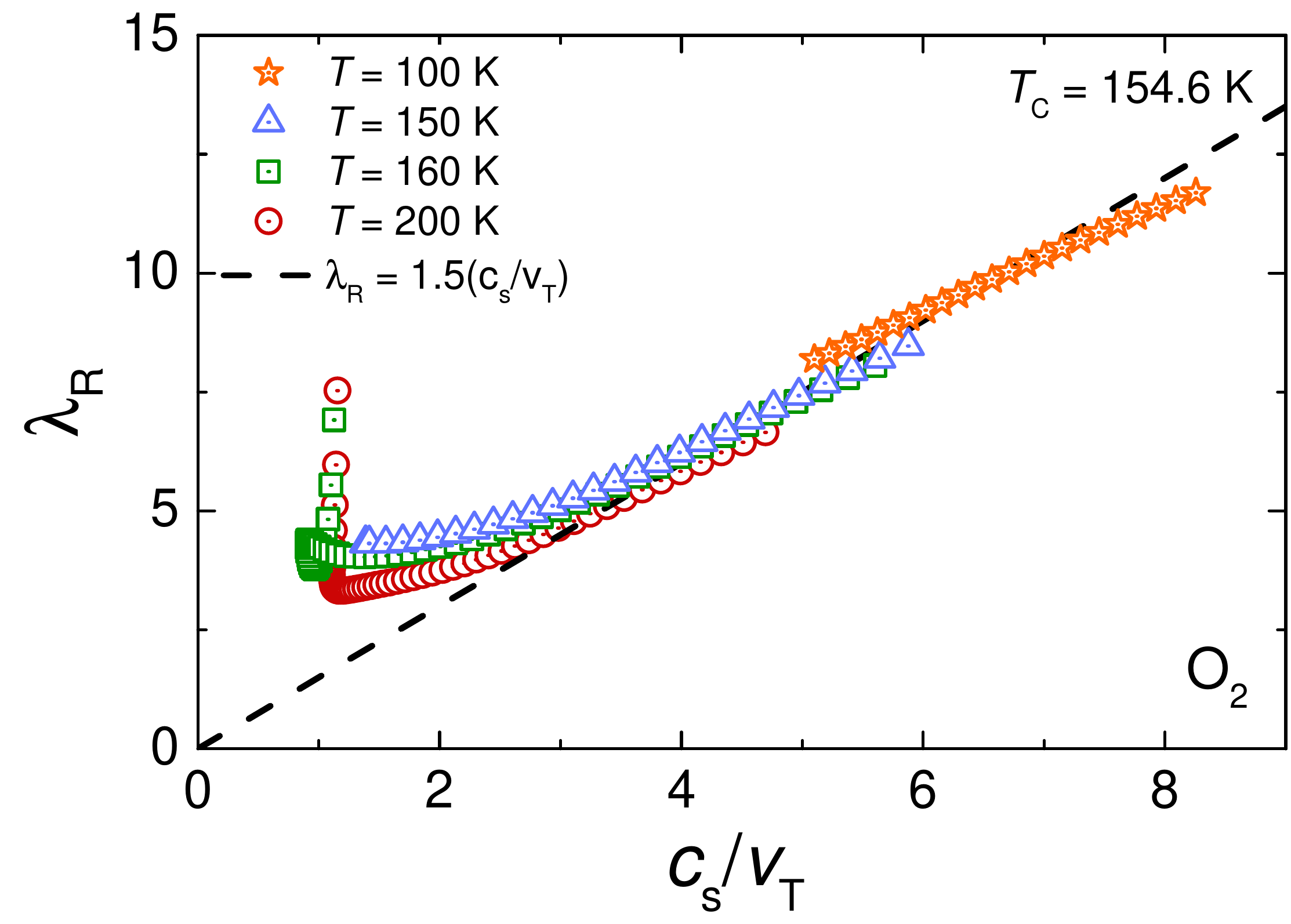}
\caption{(Color online) Macroscopically reduced thermal conductivity coefficient of liquid oxygen as a function of the reduced sound velocity. Symbols denote the values from REFPROP 10.0 along two supercritical and two subcritical isotherms (see label). The dashed line corresponds to linear correlations suggested by Bridgman's formula with the proportionality coefficient $1.5$.}
\label{Fig5}
\end{figure}

The results are plotted in Figs.~\ref{Fig4} and \ref{Fig5}. The emerging picture is to some extent similar to that in monoatomic fluids. Approximately linear correlation is preserved for $c_{\rm s}/v_{\rm T}\gtrsim 3$. The coefficients of proportionality ($\simeq 1.4$ for N$_2$ and $\simeq 1.5$ for O$_2$) are however somewhat higher than those for model and monoatomic fluids. This is likely related to a higher specific heat due to additional degrees of freedom. The values of $\lambda_{\rm R}$ at the minima on the other hand are not severely affected by this.      

\subsection{Triatomic and polyatomic molecules}

Here the thermal conductivity and sound velocity data of four important molecular liquids such as water, carbon dioxide, methane, and ethane are presented. For individual models for the EoS  and the thermal conductivity coefficient that have been implemented in REFPROP 10.0 see Refs.~\cite{Wagner2002,Huber2012} (H$_2$O), \cite{Span1996,Huber2016} (CO$_2$), \cite{Setzmann1991,Friend1989} (CH$_4$), and \cite{Bucker2006,Friend1991} (C$_2$H). For each molecular fluid we consider three supercritical and one subcritical isotherm.

\begin{figure}
\includegraphics[width=8cm]{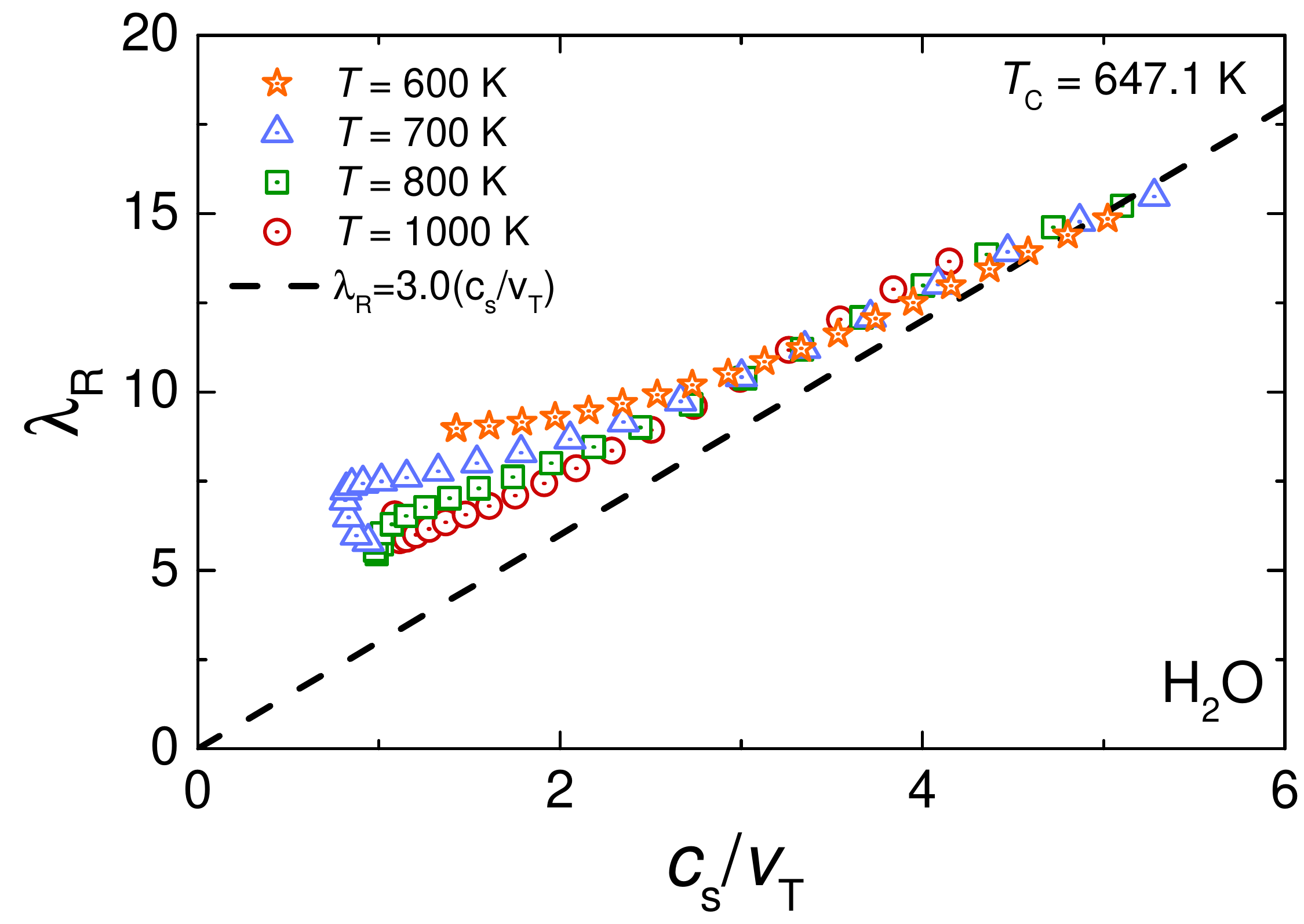}
\caption{(Color online) Macroscopically reduced thermal conductivity coefficient of water as a function of the reduced sound velocity. Symbols denote the values from REFPROP 10.0 along three supercritical and one subcritical isotherm (see label). The dashed line corresponds to linear correlations suggested by Bridgman's formula with the proportionality coefficient $3.0$.}
\label{Fig6}
\end{figure}

\begin{figure}
\includegraphics[width=8cm]{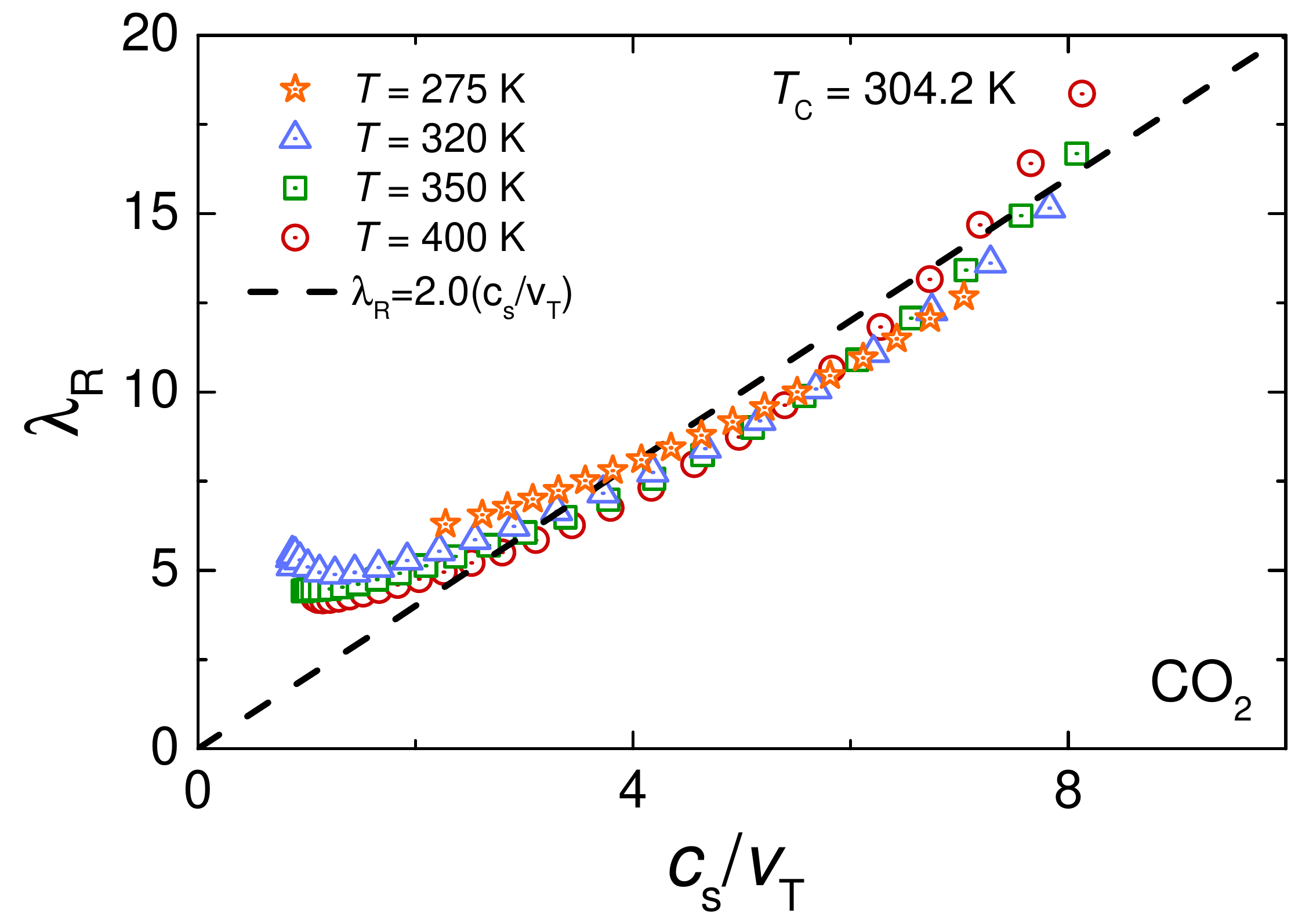}
\caption{(Color online) Macroscopically reduced thermal conductivity coefficient of carbon dioxide as a function of the reduced sound velocity. Symbols denote the values from REFPROP 10.0 along three supercritical and one subcritical isotherm (see label). The dashed line corresponds to linear correlations suggested by Bridgman's formula with the proportionality coefficient $2.0$.}
\label{Fig7}
\end{figure}

\begin{figure}
\includegraphics[width=8cm]{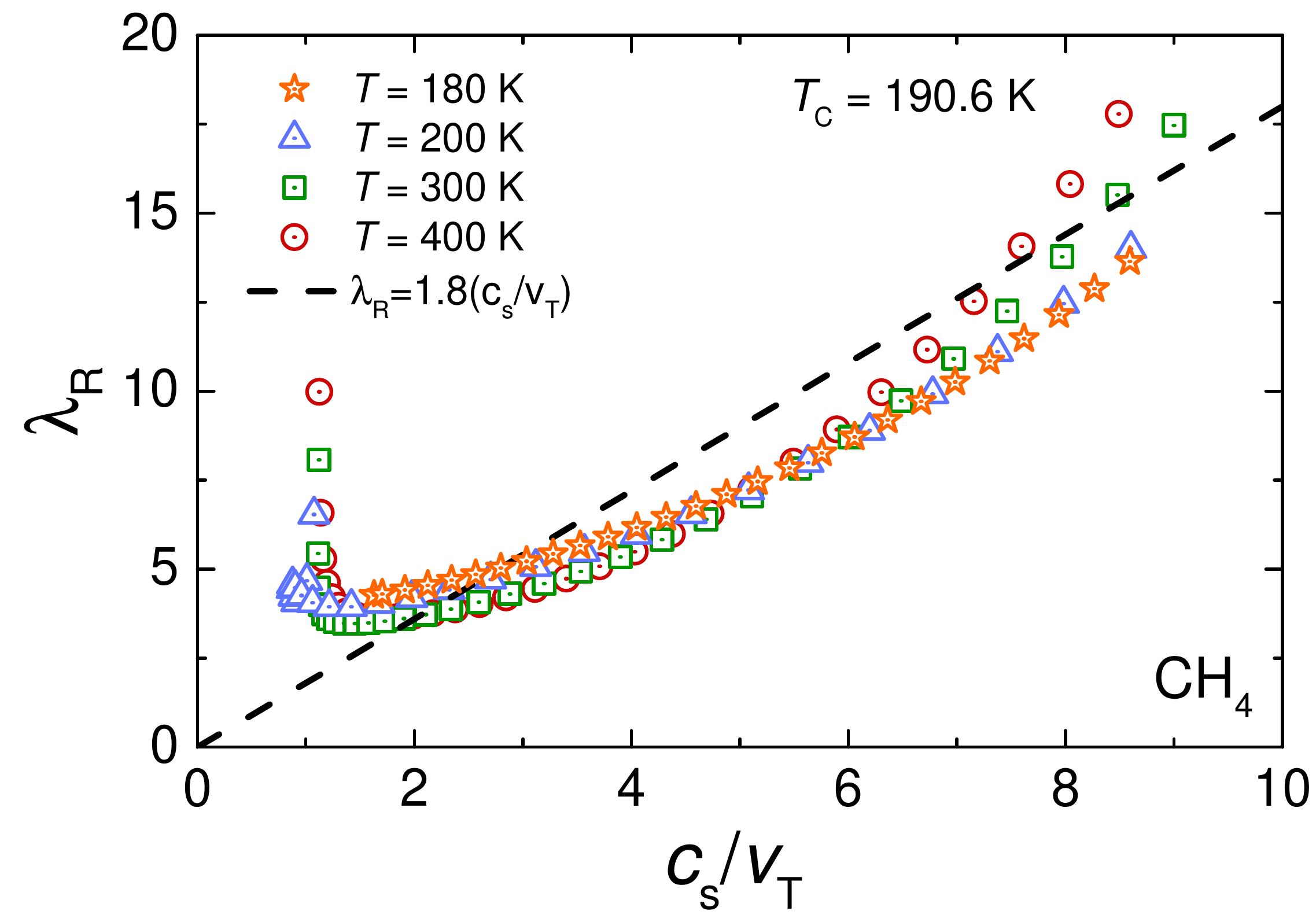}
\caption{(Color online) Macroscopically reduced thermal conductivity coefficient of methane as a function of the reduced sound velocity. Symbols denote the values from REFPROP 10.0 along three supercritical and one subcritical isotherm (see label). The dashed line corresponds to linear correlations suggested by Bridgman's formula with the proportionality coefficient $1.8$.}
\label{Fig8}
\end{figure}

\begin{figure}
\includegraphics[width=8cm]{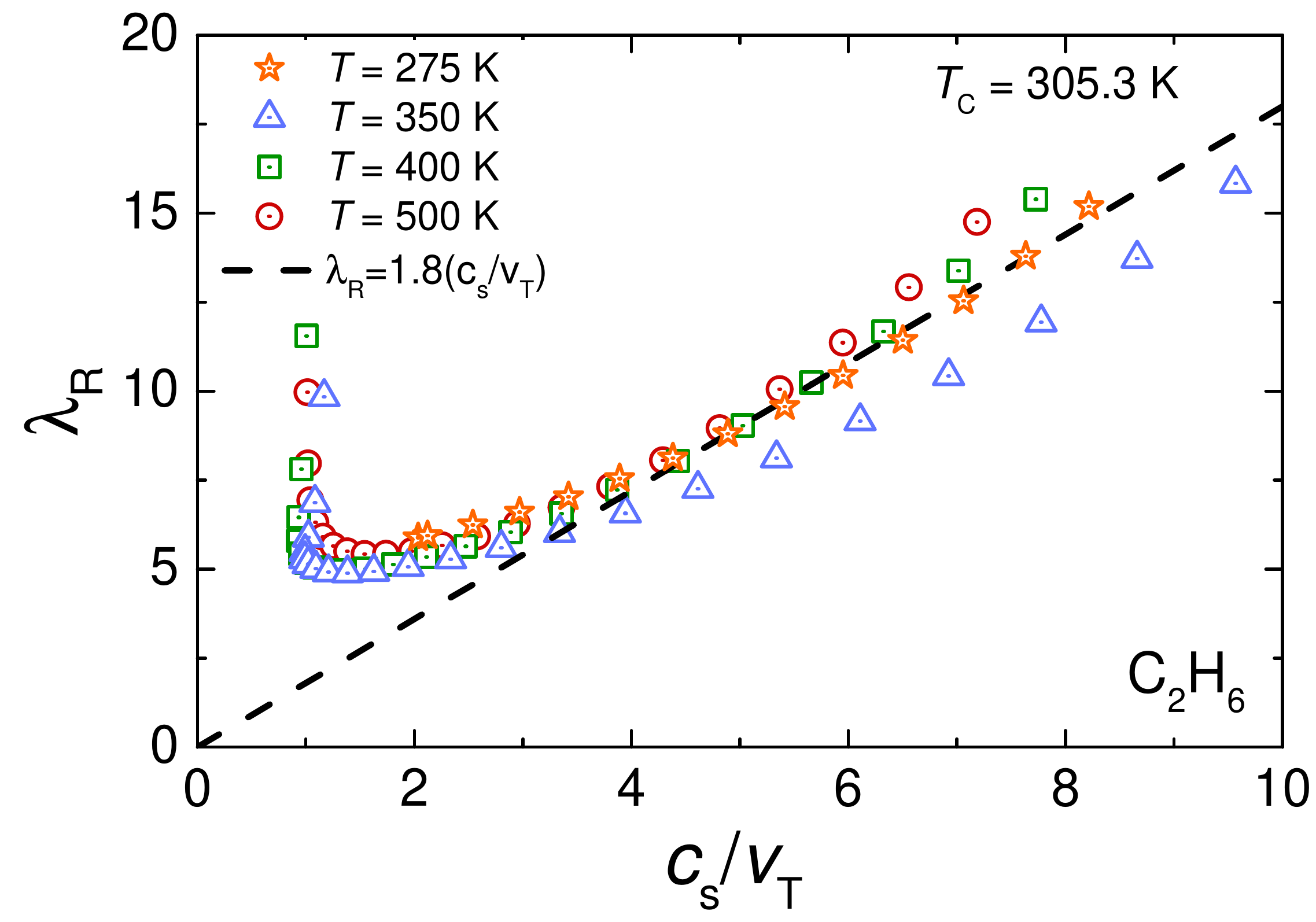}
\caption{(Color online) Macroscopically reduced thermal conductivity coefficient of ethane as a function of the reduced sound velocity. Symbols denote the values from REFPROP 10.0 along three supercritical and one subcritical isotherm (see label). The dashed line corresponds to linear correlations suggested by Bridgman's formula with the proportionality coefficient $1.8$.}
\label{Fig9}
\end{figure}

The results are shown in Figs.~\ref{Fig6} -- \ref{Fig9}. It is observed that for water a linear correlation between the thermal conductivity coefficient and the sound velocity is preserved, although an expression of the form (\ref{lin}) would be more appropriate. For the other polyatomic liquids considered the dependence of $\lambda_{\rm R}$ on $c_{\rm s}/v_{\rm T}$ has a pronounced concave shape. The linear dependence can serve only as a rather crude approximation. If nevertheless adopted, the coefficients of proportionality in Eq.~(\ref{tc1}) are $\simeq 1.8$ for methane and ethane and $\simeq 2.0$ for carbon dioxide (which is close to the original Bridgman guess). For water a larger value $\simeq 3.0$ should be used. 

Another observation is that the minima of $\lambda_{\rm R}$ have been shifted to somewhat higher values in polyatomic liquids. For carbon dioxide and methane these minima values approach $\lambda_{\rm R}^{\min}\simeq 5$. The minimum value for water is even higher. This should be again associated with an increase in specific heat due to additional degrees of freedom. 

In Figures~\ref{Fig8} and \ref{Fig9} it is observed that the quasi-universality in the dependence of $\lambda_{\rm R}$ on $c_{\rm s}/v_{\rm T}$ somewhat  degradates in methane and ethane (some dependence on the temperature becomes visible). At this point it is hard to conclude whether this observation corresponds to a real physical effect or appears as a result of uncertainties in the individual models for thermodynamics and transport.              

\section{Discussion and conclusion}\label{Conclusion}

In this paper correlations between the thermal conductivity coefficient and the sound velocity of simple model fluids, as well as real atomic and molecular liquids have been systematically investigated. The main question that has been addressed is: To which extent a simple popular formula (\ref{tc1}) proposed by Bridgman can be applicable to liquids of different molecular structure? 

It has been demonstrated that for model LJ and HS fluids, Bridgman's formula is in good agreement with the available data at sufficiently high sound velocities. The coefficient of proportionality is close to unity, in contrast to original guess by Bridgman and values assumed by others. The same is true for liquefied noble gases (as exemplified by the data for argon and krypton in this work). A linear correlation between the thermal conductivity coefficient and the speed of sound is preserved for diatomic liquids such as nitrogen and oxygen, but the coefficient of proportionality somewhat increases (to $\simeq 1.5$).      

Thus, Bridgman's formula with a proportionality coefficient $\simeq 1.2$ is expected to provide an estimate of the thermal conductivity coefficient with an error within $\simeq 10\%$ for model fluids and monoatomic liquids. For diatomic molecules the recommended coefficient is $\simeq 1.5$. Furthermore, if the thermal conductivity and the sound velocity at the melting temperature are known, the accurate determination of the proportionality coefficient is possible. If two experimental points are available, then a linear fit of the form of Eq.~(\ref{lin}) can be performed, making possible even more accurate predictions of the thermal conductivity in the entire dense liquid regime.   

For polyatomic molecular liquids the linear dependence does not appear as a very robust property. While for water, the dependence of $\lambda_{\rm R}$ on $c_{\rm s}/v_{\rm T}$ is nearly linear, for carbon dioxide, methane, and ethane the  dependence is pronouncedly concave. Linear fit would provide only a rather crude approximation in this case. The proportionality coefficient increases further. For carbon dioxide, methane and ethane the recommended coefficients are close to $\simeq 2$, as proposed originally by Bridgman. For water the proportionality coefficient reaches an even higher value of $3.0$.  
At the moment there does not appear any obvious way to relate the value of the proportionality coefficient to molecular structure or other property of a polyatomic liquid. Hence, for polyatomic liquids Bridgman's formula is less useful than for monoatomic and diatomic liquids.

One should note that there are situations, where Bridgman's formula is not applicable and should not be used. First of all, it can only be applied in the dense liquid regime at densities above the gas-like to liquid-like dynamical crossover, marked by the Frenkel line~\cite{BrazhkinPRE2012,BrazhkinPRL2013,BellJCP2020,KhrapakJCP2022}. From the evidence presented in this work, the onset of validity of Bridgman's formula corresponds to reduced sound speeds exceeding $\simeq 3 v_{\rm T}$. Additionally, sound velocity must simply exist. There are exotic systems with extremely soft interactions, where the longitudinal collective mode dispersion relation is not acoustic and the sound velocity formally diverges. An excellent example is related to one-component plasma and weakly screened Coulomb systems in the plasma-related context. The Bridgman's formula is irrelevant to such fluid systems. On the other hand, a vibrational model of heat transport has been demonstrated to work rather well in fluids with soft pairwise interactions~\cite{KhrapakPRE01_2021,KhrapakPoP2021,KhrapakPoP08_2021,
KhrapakMolecules12_2021,KhrapakPPR2023}.             

To conclude, Bridgman's formula relating the thermal conductivity coefficient and sound velocity of dense liquids appears quite useful for simple model fluids as well as real monoatomic and diatomic liquids. For polyatomic molecular liquids, on the other hand, the linear proportionality between the thermal conductivity coefficient and sound speed is not so appealing and the proportionality coefficient can vary unpredictably. Here Bridgman's formula can only be used as a rough preliminary estimate. In general, the dependence of the proportionality coefficient on molecular structure has to be properly accounted for. These findings represent an important step toward better understanding transport properties of various atomic and molecular liquids in different parameter regimes.

The author declares no conflict of interests.





\bibliography{SE_Ref}

\end{document}